\documentstyle[12pt,moriond,psfig]{article}
\begin{document}
\heading{STAR FORMATION HISTORIES OF NEARBY\\ GALAXIES AND THE
CONNECTION TO HIGH\\ REDSHIFT}

\author{Eline Tolstoy} {Space Telescope - European Coordinating
Facility, Garching bei M\"{u}nchen, Germany}

\begin{moriondabstract}

It is an obvious statement that all the galaxies we see today in and around 
our Local Group have been forming and evolving for a significant fraction 
of the age of the Universe.  It is not a great leap of logic to further 
state that the manner in which they have formed and evolved must be fairly 
representative of these processes in general.  Unless of course we would 
like to assume that our local region of space is in some way peculiar for 
which there is no evidence.  In other words, if we are able to determine 
accurate star formation histories for the nearby galaxies back to the ages 
of the oldest globular clusters then we will also obtain a representative 
picture of how galaxies have evolved from the earliest times, and predict 
what nearby galaxies looked like at intermediate and high redshifts.

Deep, precision, multi-colour photometry of resolved stellar populations in 
external galaxies can uniquely determine the star formation histories of 
nearby galaxies going back many Gyrs.  {\it Hubble Space Telescope} and 
high quality ground based imaging have recently resulted in dramatic 
Colour-Magnitude Diagrams of the faint old resolved stellar populations in 
nearby galaxies.  Although data deep enough to unambiguously trace back to 
the very oldest populations does not yet exist (except for a few small 
dSphs in the Galaxy Halo), these preliminary studies make the potential for 
deeper data for a range of galaxy types look very exciting.  For example, 
recent results on the very low metallicity dwarf irregular galaxy Leo~A 
suggests we have found a predominantly, if not entirely, young galaxy, less 
than 2~Gyr old, in the Local Group.  Nearby faint star CMDs thus provide an 
important and independent method to confirm the high redshift galaxy survey 
predictions of galactic evolution.

\end{moriondabstract}
\newpage
\section{Colour-Magnitude Diagram Analysis}

Stellar evolution theory provides a number of clear predictions, based on 
relatively well understood physics, of features expected in 
Colour-Magnitude Diagrams (CMD) for different age and metallicity stellar 
populations (see Figure~1).  There are a number of clear indicators of 
varying star formation rates ({\it sfr}) at different times which can be 
combined to obtain a very accurate picture of the entire star formation 
history (SFH) of a galaxy.

\subsection{Main Sequence Turnoffs (MSTOs)}

If we can obtain deep enough exposures of the resolved stellar populations 
in nearby galaxies we can obtain the {\it unambiguous age information that 
comes from the luminosity of MSTOs}.  Along the Main Sequence (MS) itself 
different age populations overlie each other completely making the 
interpretation of the MS luminosity function complex, especially for older 
populations.  However the MSTOs do not overlap each other like this and 
hence provide the most direct, accurate information about the SFH of a 
galaxy.  MSTOs can clearly distinguish between bursting star formation and 
quiescent star formation, {\it e.g.}  \cite{hk98}.  The age resolution that is 
possible does vary, becoming coarser going back in time.  Our ability to 
disentangle the variations in {\it sfr} depends upon the the intensity of 
the past variations and how long ago they occured and which filters are 
used for observation.  For ages less than about 1.5~Gyr it is possible to 
have detailed resolution on the 10$-$100~Myr time scales.  As can be seen in 
Figure~1, the ages begin to crowd together for older populations.  Beyond 
about 8~Gyr ago the age resolution with optimum data is on the scale of 
roughly a Gyr.  This of course means that a very short high intensity burst 
of star formation in this period will be ``spread out '' over a Gyr time 
period and so seem less intense and longer lasting.  As will be described 
in the following sections, however, there are other indicators in a CMD 
which help to narrow down the range of possible SFHs.

\hskip -1.5cm
\psfig{figure=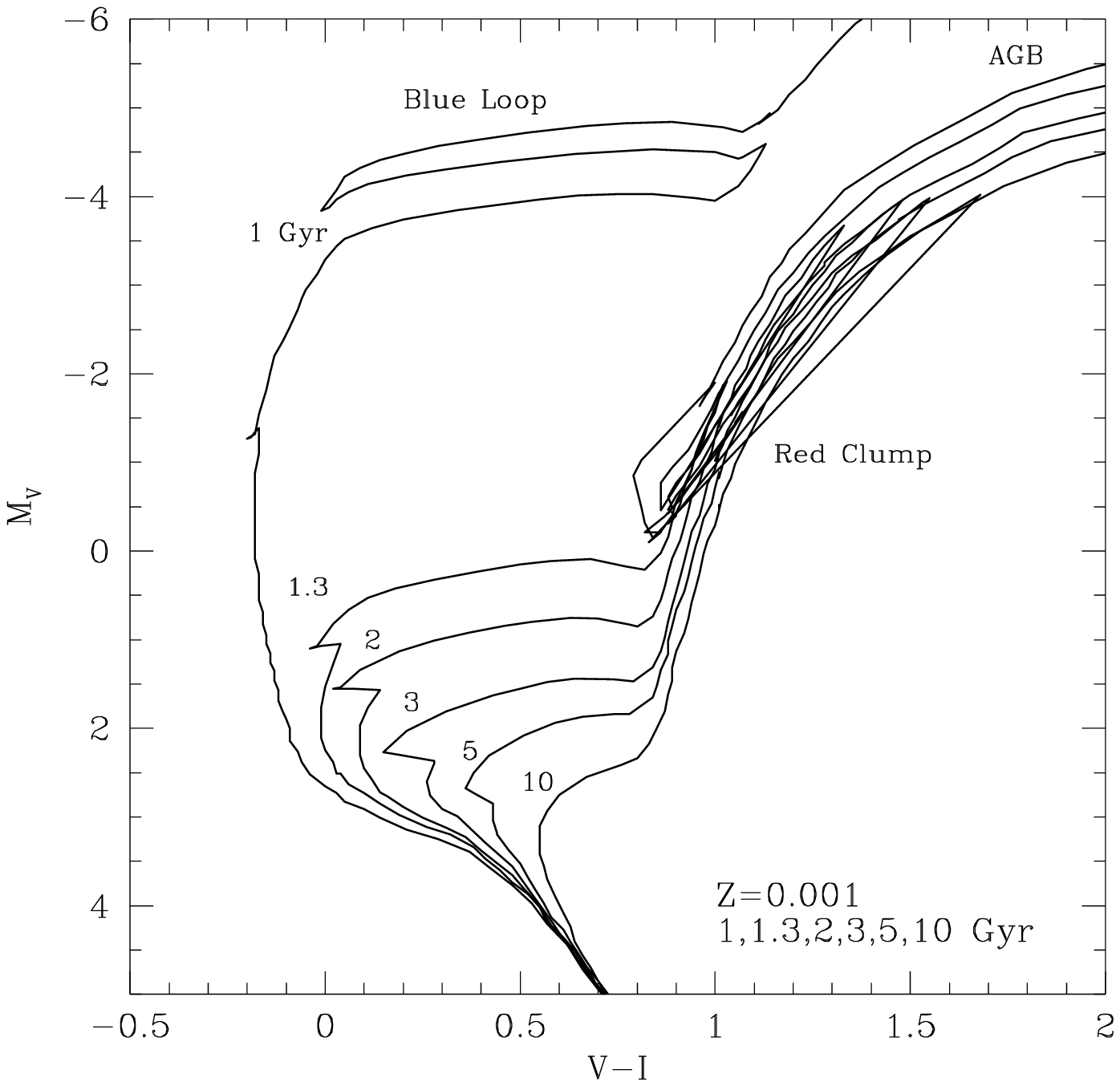,height=10cm,width=10cm}
\vskip -7.1cm
\hskip 7.cm
\begin{picture}(100,30)(0,0)
\put(0,0){\parbox{8.5cm}{\small Figure~1: Isochrones for a single
metallicity (Z=0.001) and a range of ages, as marked in Gyr
\cite{b94}, at the MSTOs.  Isochrones were designed for single age
globular cluster populations and are best avoided in the
interpretation of composite populations, which can best be modeled
using Monte-Carlo techniques ({\it e.g.} \cite{et96}).
}}
\end{picture}
\vskip4.cm

\subsection{The Core-Helium Burning Blue Loop Stars (BLs)}

Stars of certain metallicity and mass go on, what are elegantly refered to 
as, ``Blue Loop Excursions'' after they ignite He in their core.  Stars in 
the BL phase are several magnitudes brighter than when on the MS.  They 
thus provide a more luminous opportunity to accurately determine the age 
and metallicity of the young stellar population (in the range, 
$\lsim$~1~Gyr old) in nearby low metallicity galaxies
\cite{dp, dp1, dp2}.  The shape and mass at which these ``loops'' are
seen in a CMD are a strong function of metallicity and age, and the
luminosity of a BL star is fixed for a given age. Subsequent
generations of BL stars do not overlie each other as they do on the
MS.  The lower the metallicity of the galaxy the older will be the
oldest BLs and the further back in time an accurate SFH can easily be
determined. BL stars are brighter than MS stars of the same mass,
often by more than a magnitude.

\subsection{The Red Giant Branch (RGB)}

The RGB is a very bright evolved phase of stellar evolution, where the star 
is burning H in a shell around its He core.  For a given metallicity the 
RGB red and blue limits are given by the young and old limits 
(respectively) of the stars populating it (for ages $\gsim$1~Gyr).  As a 
stellar population ages the RGB moves to the red, for constant metallicity, 
the blue edge is determined by the age of the oldest stars.  However 
increasing the metallicity of a stellar population will also produce 
exactly the same effect as aging, and also makes the RGB redder.  This is 
the (in)famous age-metallicity degeneracy problem.  The result is that if 
there is metallicity evolution within a galaxy, it impossible to uniquely 
disentangle effects due to age and metallicity on the basis of the optical 
colours of the RGB alone.

\vskip.5cm
\hskip -.7cm
\psfig{figure=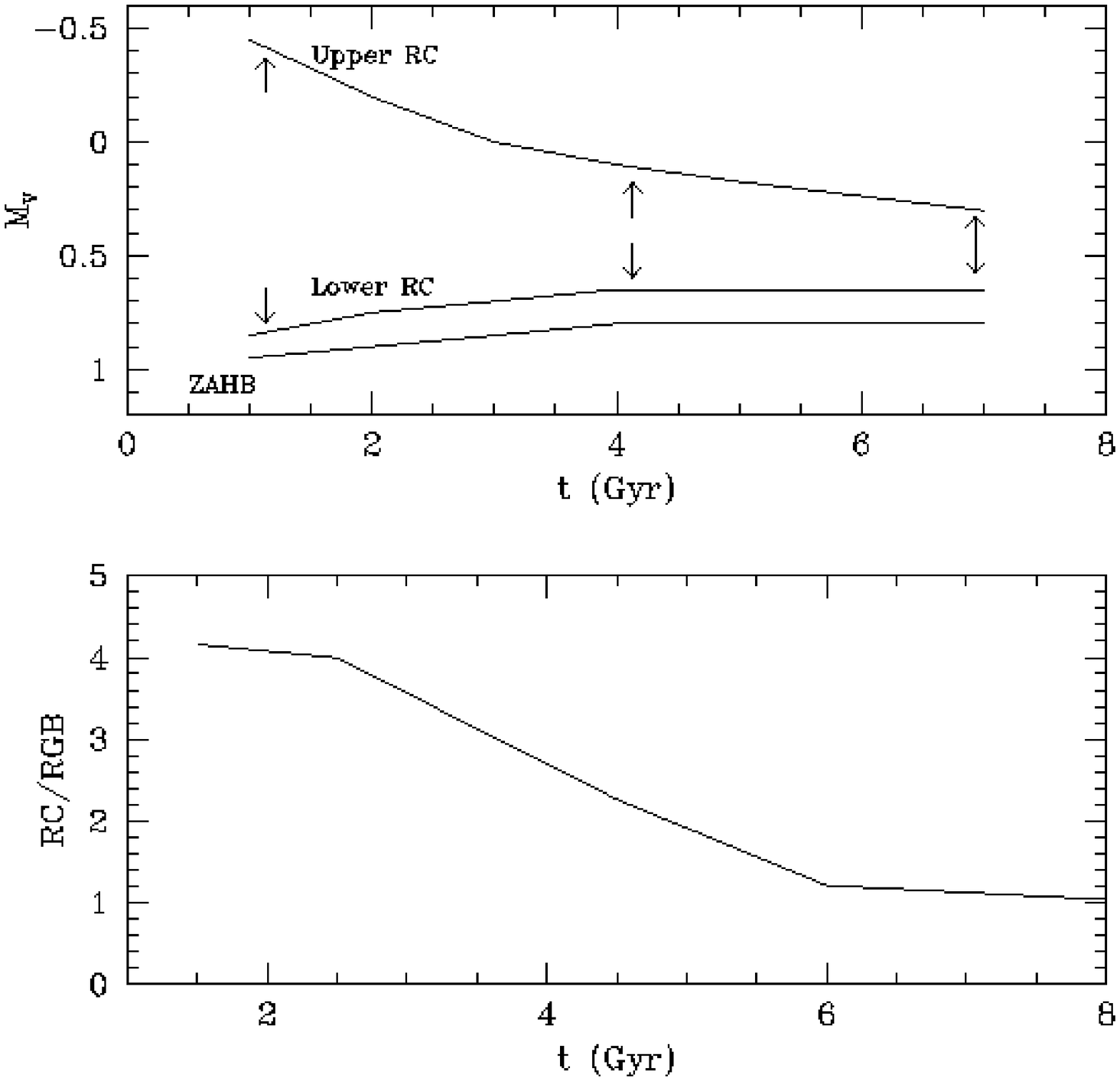,height=8cm,width=8cm}
\vskip -6.2cm
\hskip 7.8cm
\begin{picture}(100,60)(0,0)
\put(0,0){\parbox{8.5cm}{\small Figure~2: In the top panel are plotted
the results of Caputo, Castellani \& Degl'Innocenti \cite{ccd} for the 
variation in the {\it extent} in M$_V$ magnitude of a RC with age, for a 
metallicity of Z=0.0004.  We plot the magnitude of the upper and lower edge 
of the RC versus age, in Gyr.  We can thus clearly see that this extent is 
strong function of the age of the stellar population.  Also plotted is 
M$_V$ of the zero age HB against age.

In the bottom panel are plotted the results of running a series of 
Monte-Carlo simulations \cite{et96} using stellar evolution models at 
Z=0.0004 \cite{f94} and counting the number of RC and RGB stars in the same 
part of the diagram, and thus we determine the expected ratio of RC/RGB 
stars versus age.
}}
\end{picture}
\vskip4.3cm

\subsection{The Red Clump/Horizontal Branch (RC/HB)}

Red Clump (RC) stars and their lower mass cousins, Horizontal Branch (HB) 
stars are core helium-burning stars, and their luminosity varies depending 
upon age, metallicity and mass loss \cite{ccd}.  The extent in luminosity 
of the RC can be used to estimate the age of the population that produced 
it \cite{ccd}, as shown in the upper panel of Figure~2.  This age 
measure is {\it independent of absolute magnitude and hence distance}, and 
indeed these properties can be used to determine an accurate distance 
measure on the basis of the RC \cite{c98}.

The classical RC and RGB appear in a population at about the same time 
($\sim$ 0.9--1.5 Gyr, depending on model details), where the RGB are the 
progenitors of the RC stars.  The lifetime of a star on the RGB, t$_{RGB}$, 
is a strongly decreasing function of M$_{star}$, but the lifetime in the 
RC, t$_{RC}$ is roughly constant.  Hence the ratio, t$_{RC}$ / t$_{RGB}$, 
is a decreasing function of the age of the dominant stellar population in a 
galaxy, and the ratio of the numbers of stars in the RC, and the HB to the 
number of RGB is sensitive to the SFH of the galaxy \cite{et98, han97}.  
Thus, the higher the ratio, N(RC)/N(RGB), the younger the dominant stellar 
population in a galaxy, as shown in the lower panel of Figure~2.

The presence of a large HB population on the other hand (high N(HB)/N(RGB) 
or even N(HB)/N(MS), is caused by a predominantly much older ($>$10~Gyr) 
stellar population in a galaxy.  The HB is the brightest indicator of very 
lowest mass (hence oldest) stellar populations in a galaxy.

\subsection{The Extended Asymptotic Giant Branch (EAGB)}

The temperature and colour of the EAGB stars in a galaxy are determined by 
the age and metallicity of the population they represent (see Figure~3).  
However there remain a number of uncertainties in the comparison between 
the models and the data \cite{g94, L98}.  It is very important that more 
work is done to enable a better calibration of these very bright indicators 
of past star formation events.  In Figure~3 theoretical EAGB isochrones 
\cite{b94} are overlaid on the HST CMD of a post-starburst BCD galaxy and we 
can see that a large population of EAGB stars is a bright indicator of a 
past high {\it sfr}, and the luminosity spread depends upon metallicity and 
the age of the {\it sfr}.

\vskip -7.5cm
\hskip -1.cm
\psfig{figure=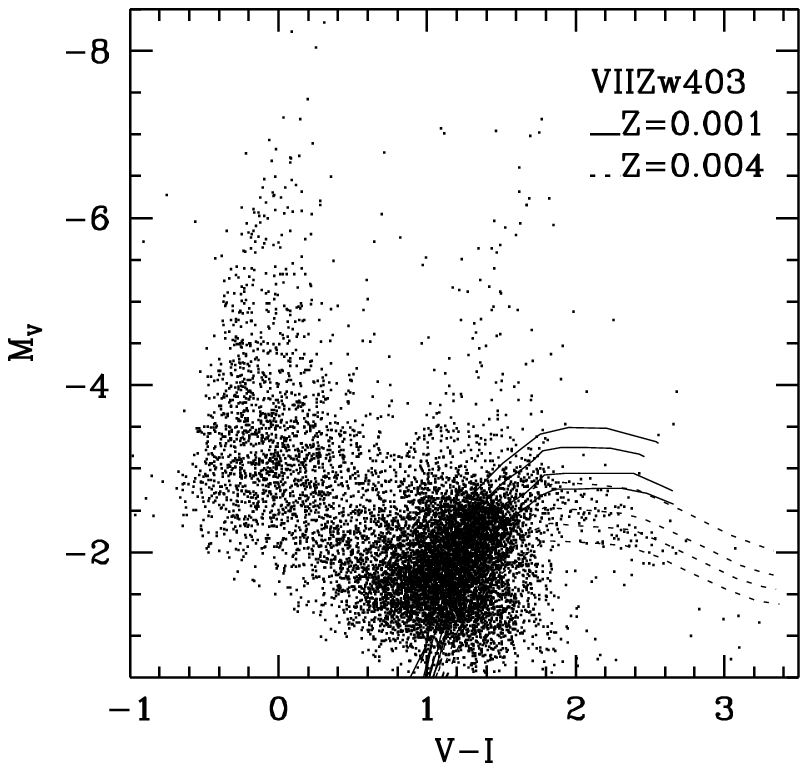,height=15cm,width=15cm}
\vskip -5.1cm
\hskip 6.8cm
\begin{picture}(100,30)(0,0)
\put(0,0){\parbox{8.5cm}{\small Figure~3: EAGB isochrones \cite{b94}
for metallicities, Z=0.001 and Z=0.004, are shown superposed on the 
observed CMD of VII~Zw403 \cite{L98}.  For each metallicity the isochrones 
are for populations of ages 1.3, 2, 3, and 5 Gyrs, with the youngest 
isochrone being the brightest.  This shows the potential discriminant 
between the age and metallicity of older populations, if the models could 
be better calibrated to a known SFH, {\it e.g.}  for a nearby EAGB rich system 
like NGC~6822 where old MSTOs are observable.
}}
\end{picture}
\vskip 3.3cm

\subsection{Distance, Extinction \& Metallicity}

The accurate interpretation of the indicators described above depends 
critically upon having reliable estimates for the distance, the extinction 
and the metallicity of a galaxy.  Ideally we would also like to know if the 
extinction is patchy and on what scale, and what has been the evolution of 
the metallicity of the stellar population with time.

The basic properties of distance, extinction and metallicity can be 
determined independently to the CMD, and they must be consistent with the 
findings in a CMD.  These three basic parameters, in conjunction with 
observational errors and incompleteness make the most significant impact on 
the properties of the CMD and hence the final SFH model \cite{et96}.  There 
are a number of difficulties in accurately determining these basic 
properties but they can be resolved with careful observation and analysis 
techniques.

\subsubsection{The Distance} is the most crucial parameter for accurate
analysis of a CMD, partly because it can easily be wrong by many orders of 
magnitude (for example the young, red supergiants can be mistaken for the 
RGB, if the observed CMD isn't deep enough to confirm the identification, 
i.e.  by detecting a RC or HB).  If the distance to a galaxy is incorrect 
this will result in the masses of individual stars being wrongly 
determined, and hence the age of the different populations will be wrong.  
Distances are most accurately determined by primary distance indicators 
({\it e.g.}  RR~Lyr or Cepheid variable stars), but there is also useful 
information in the tip of the RGB \cite{lfm}; the RC \cite{c98}; BLs 
\cite{et98}.  To be sure of the distance to small faint galaxies it is 
necessary to have a CMD which goes deep enough to extend below the RC/HB.

\subsubsection{The Extinction}, both internal to a galaxy and between us 
and a galaxy can affect the accurate analysis of a CMD.  If the extinction 
is incorrectly determined it will have the same effect as a distance error, 
and hence effect the reliability of the SFH models.  Local HI maps in 
conjunction with Infra-Red ({\it e.g.}  IRAS) 100$\mu$m maps can provide an 
accurate picture of how much extinction can be expected in any given 
direction in the sky \cite{g98}.

\subsubsection{Metallicity:} When a galaxy makes stars, then the detritus 
of this process ({\it e.g.}, from SN explosions and stellar winds) make it 
unlikely that the galaxy can avoid metallicity evolution altogether 
\cite{els}.  However, there is no concrete observational evidence that this 
is true, although abundance ratios of different elements do give us model 
dependent suggestions \cite{p94}.  In the disc of our Galaxy, for example, 
it was recently shown that, although there is a general trend in 
metallicity evolution with time, the scatter is always large \cite{ed}.  We 
do not understand in detail how stars interact with their surrounding ISM, 
and thus how current star formation feeds the metal enrichment of future 
generations.  Looking at recent results of absorption line studies of Zinc 
abundances at cosmological distances ($z = 0.7 - 3.4$) there is evidence 
for a shallow evolution of metallicity in galaxies over this long redshift 
range, but there is also a large scatter in values at any time \cite{p97}.  
Absorption line studies of these species provides arguably the most 
reliable estimator of the metallicity of the {\it gas} in a galaxy.  If 
suitable background continuum sources could be found behind nearby galaxies 
this would dramatically improve our understanding of how the ISM in 
different galaxies evolves and is affected by the proximity of current star 
formation.

Accounting for metallicity evolution in a CMD is difficult.  It is 
impossible to determine a unique model based solely on the RGB because of 
age-metallicity degeneracy.  However, if metallicity evolution is neglected 
in a CMD model then the best model for that galaxy will typically be 
younger than if metallicity evolution were included \cite{et98}.

Understanding the details of metallicity evolution in galaxies is one of 
the most critical areas for further study if we are to develop an accurate 
understanding of galaxy evolution.

\section{Recent Results from {\sl HST} Observations}

An HST program was initiated by Skillman \cite{s98}, using four orbits of 
telescope time per galaxy, in three filters (effectively B, V and I), to 
study a sample of four nearby dwarf irregular (dI) galaxies.  The initial 
sample consists of: Sextans~A, Pegasus, Leo~A \& GR~8.  The results have 
been dramatic and illustrate the tremendous advances possible, even with 
short exposures, when crowding has been virtually illuminated \cite{dp, 
dp1, dp2, dp3, g98, s98, et98}.

\subsection{Leo A: A Predominantly Young Galaxy within the Local Group}

The unprecedented detail of the WFPC2 CMDs of the resolved stellar 
population of Leo~A has resulted in an improved distance determination and 
an accurate SFH for this extremely metal-poor Local Group (LG) dI galaxy 
\cite{et98}.  From the position of the RC, the BLs and the tip of the RGB, 
a distance modulus, m$-$M=24.2$\pm$0.2, or 690 $\pm$ 60 kpc, was obtained 
which places Leo~A firmly within the LG.  The interpretation of these 
features in the WFPC2 CMDs at this new distance based upon extremely low 
metallicity (Z=0.0004) theoretical stellar evolution models suggests that 
this galaxy is predominantly young, {\it i.e.} $<$~2~Gyr old.  A major 
episode of star formation 900$-$1500~Gyr ago can explain the RC luminosity 
and the ratio of N(RC)/N(RGB) stars as well as consistency with the the 
number of anomalous Cepheid variable stars seen in this galaxy.  The 
presence of an older, underlying globular cluster age stellar population 
could not be ruled out with these data.  However, using the currently 
available stellar evolution models, it would appear that such an older 
population is limited to no more than 10\% of the total star formation to 
have occured in this galaxy.  The HST CMDs and the modeling results are 
shown below in Figure~4.

\hskip -0.5cm \centerline{
\psfig{figure=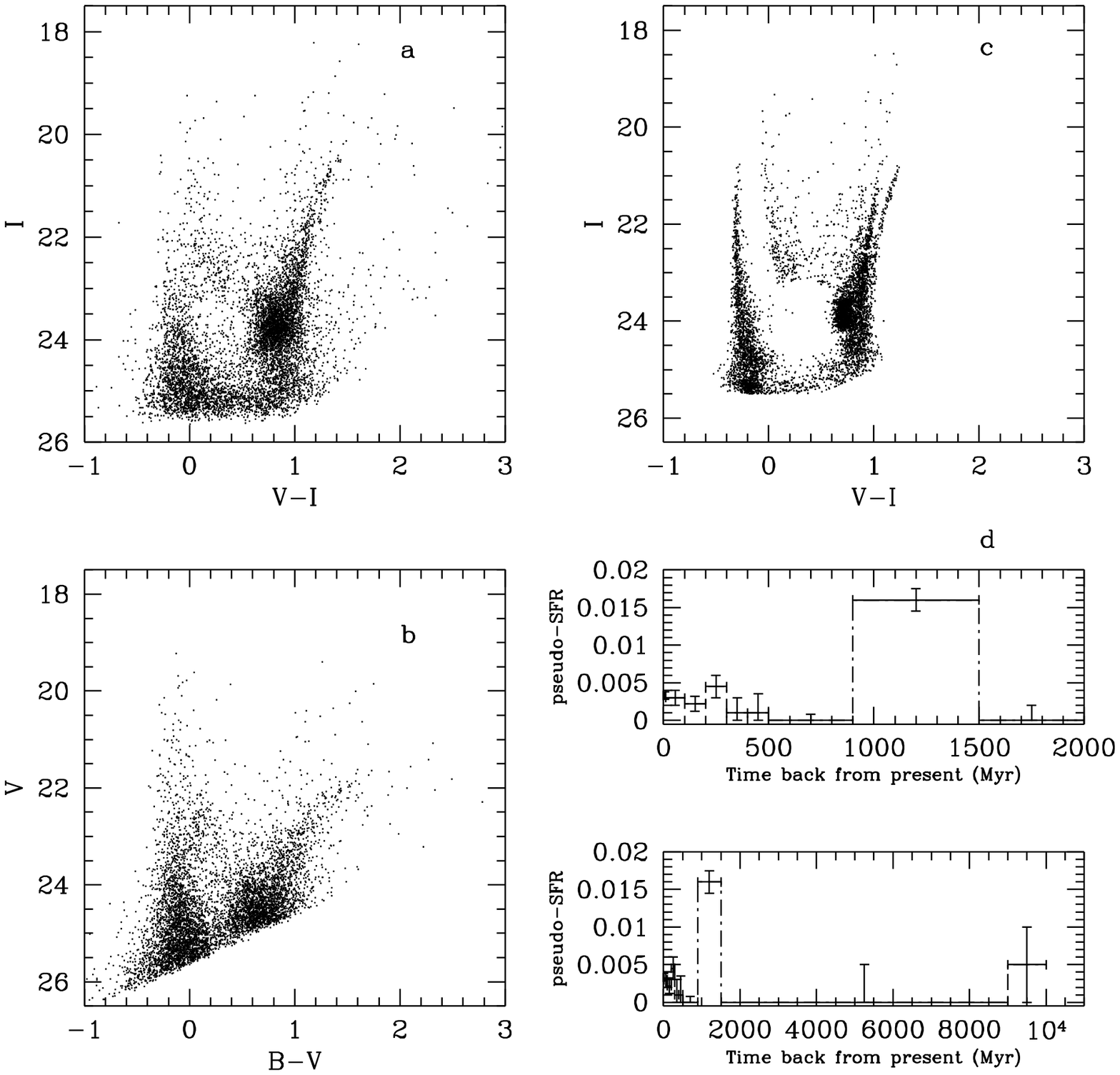,height=15cm,width=15cm}} \vskip0.3cm
\begin{picture}(300,30)(0,0)
\put(0,0){\parbox{15cm}{\small Figure~4: Here we show the results for
the analysis of the HST/WFPC2 data \cite{et98}.  In a.  is the V$-$I, I HST 
CMD for Leo~A, 1 orbit exposure time per filter.  In b.  is the B$-$V, V 
HST CMD for Leo~A, 2 orbits in B (F439W).  In c.  is the best match 
Monte-Carlo simulation model (in V$-$I, I) found for these data convolved 
with the theoretical measurement error distribution \cite{ts}, and in d.  
is the SFH that created the model CMD which best matches these data.  See 
Tolstoy {\it et al.}  \cite{et98} for more details.
}}
\end{picture}
\vskip 1.5cm

\subsection{Pegasus: A Not so Young Galaxy?}

The resolved stellar population of the Pegasus dI galaxy reveals quite a 
different SFH to Leo~A \cite{g98}.  A young ($<$ 0.5 Gyr), and weak MS 
stellar component is also present.  In Pegasus however, the distinctive BLs 
are not visible in the data.  This may be due to spatial variations in the 
internal extinction properties of Pegasus which effectively smear out this 
feature which would already be weak in a galaxy with such a small young 
population.  The colours of the MS also suggest an unexpectedly large 
foreground extinction of A$_V$ = 0.47 mag.  The width in colour of the RGB 
implies a range of stellar ages and/or metallicities.  A small number of 
EAGB stars are found beyond the RGB tip and in WIYN ground based imaging 
\cite{g98} and near the faint limits of the HST data is a populous RC.  
Fitting a self-consistent stellar population model based on the Z$=$0.001 
Geneva stellar evolution tracks yielded a revised distance of 760 kpc.  The 
numbers of MS and BL stars require that the {\it sfr} was higher in the 
recent past, by a factor of $\sim$10 about 2~Gyr ago, assuming no 
metallicity evolution.  Unique results cannot be obtained for the SFH over 
longer time baselines without better information on stellar metallicities 
and deeper photometry.  Even at its peak of star forming activity, Pegasus 
most likely remained relatively dim with M$_V \sim -$14.  The HST CMDs and 
the modeling results for Pegasus are shown in Figure~5 below.

\hskip -0.5cm \centerline{
\psfig{figure=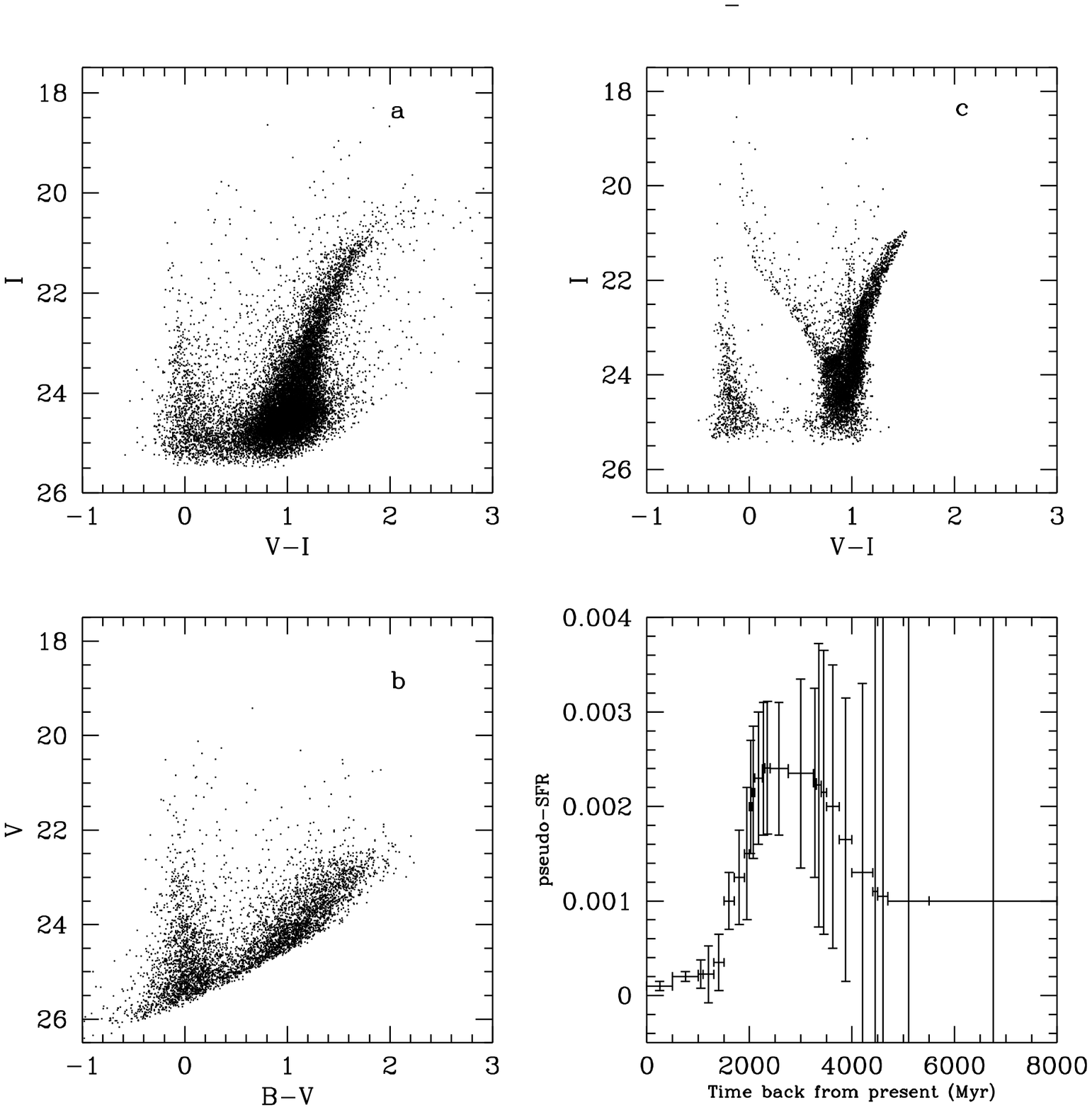,height=15cm,width=15cm}} \vskip0.3cm
\begin{picture}(300,30)(0,0)
\put(0,0){\parbox{15cm}{\small Figure~5: Here we show the results for
the analysis of the HST/WFPC2 data \cite{g98}.  In a.  is the V$-$I, I HST 
CMD for Pegasus, 1 orbit exposure time per filter.  In b.  is the B$-$V, V 
HST CMD for Pegasus, 2 orbits in B (F439W).  In c.  is the best match 
Monte-Carlo simulation model \cite{ts}, in V$-$I, I, found for these data 
({\it excluding the RC}) and convolved with the theoretical measurement 
error distribution, and in d.  is the SFH that created the model CMD which 
best matches the data.  
See Gallagher {\it et al.}  \cite{g98} for more details.
}}
\end{picture}
\vskip 1.5cm

\subsection{Sextans~A \& GR~8: Detailed Star Formation Patterns}

Sextans~A and GR~8 are the furthest away in the Skillman sample (at 1.4 and 
1.6 Mpc respectively) and so these data do not allow us to consider SFHs 
beyond $\sim$800~Myr ago, however they do provide detailed information 
about how star formation has varied spatially across these small galaxies 
on time scales $<$~1~Gyr using the MS and BL stars.  There is insufficient 
information to obtain unique information from the RGB about the older 
populations from these data.  The Dohm-Palmer contribution to this volume 
\cite{dp} discusses these results in detail, and see \cite{dp1, dp2, dp3}.

\subsection{VII~Zw 403: A Post-Starburst Blue Compact Dwarf Galaxy}

Another interesting new result from HST comes from a study of the nearest 
by Blue Compact Dwarf (BCD) Galaxy, VII~Zw 403 by 
Lynds {\it et al.}  \cite{L98}.  
The (V$-$I, M$_V$) HST CMD is shown in Figure~3.  Another study of the same 
HST observations of VII~Zw 403 is presented by Schulte-Ladbeck, in this 
volume.  This galaxy is at a distance of $\sim$5~Mpc, and thus it of 
similar resolution to ground based observations of nearby dI type galaxies, 
such as NGC~6822 \cite{g94}.  The similarity between Figure~3 and the 
ground-based CMD of NGC~6822 of Gallart {\it et al.} \cite{g94} is quite 
startling.  Clearly this is a similarity which needs further study.  
NGC~6822 is also close enough to calibrate the EAGB versus SFH via the 
detection of older MSTOs.

\section{The Connection to High Redshift}

Star-forming, dI galaxies represent the largest fraction by number of 
galaxies in the LG, and it is clear from deep imaging surveys that this 
number count dominance appears to {\it increase} throughout the Universe 
with lookback time \cite{e97}.  The large numbers of ``Faint Blue 
Galaxies'' (FBG) found in deep imaging-redshift surveys appear to be 
predominantly intermediate redshift ($z<1$, or a look-back time out to 
roughly half a Hubble time), intrinsically {\it small} late type galaxies, 
undergoing strong bursts of star formation \cite{bf96}.  Thus we can assume 
that the dIs we see in the LG ({\it e.g.}  
Leo~A, Pegasus, Sextans~A, etc.) are a 
cosmologically important population of galaxies which can be used to trace 
the evolutionary changes in the {\it sfr} of the Universe with redshift.  
The ``Madau-diagram'' \cite{madau} uses the results of redshift surveys to 
plot the SFH of the Universe against redshift.  It predicts that most of 
the stars that have formed in the Universe have done so at redshifts, {\it 
z} $\sim 1 - 2$.  If it is correct, then the MSTOs from the most active 
period of star formation in the Universe will be easily visible as 
7$-$9~Gyr old MSTOs in the galaxies of the LG \cite{Rich}.  
Determining accurate SFHs 
for all the galaxies in the local Universe using CMD analysis provides an 
alternate route to and thus check upon the Madau-diagram.

Recent detailed CMDs of several nearby galaxies and self-consistent grids 
of theoretical stellar evolution models have transformed our understanding 
of galactic SFHs.  Both of the dIs we have looked at, and for which we 
detected a RC, (Leo~A and Pegasus) agree that the {\it sfr} was higher in 
the past, although the peak in the sfr has occured at relatively recent 
times as defined by Madau-diagram (the peaks occur at z=0.1$-$0.2, within the 
first bin).  The Mateo review of {\it all} LG dwarf galaxies \cite{mario} 
and studies of M31 and our Galaxy \cite{Ren}, 
on the other hand, suggest that the LG 
had its most significant peak in star formation $>$10~Gyr ago (i.e at 
z~$>$~3), the epoch of halo formation.  Many galaxies contain large
numbers of RR~Lyr variables (or HB) and/or globular clusters which can only 
come from a
significant older population.  It is possible that dI galaxies have quite 
different SFHs to the more massive galaxies.  Thus although the small dI 
galaxies in the LG have been having short, often intense, bursts of star 
formation in comparatively recent times this is not representative of the 
majority of the star formation in

\vskip -7cm
\centerline{
\psfig{figure=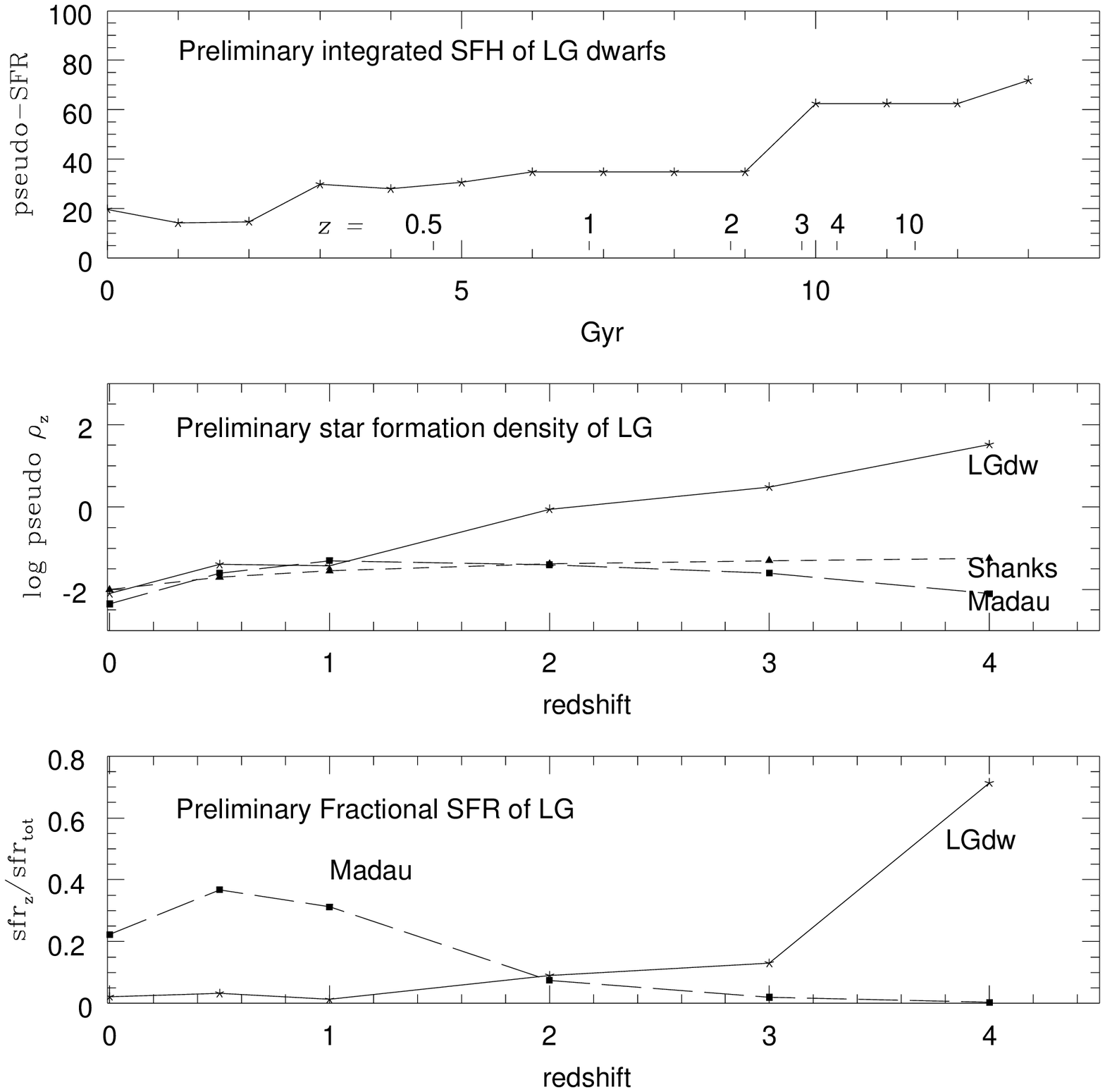,height=10cm,width=10cm}} 
\vskip 1.7cm
\begin{picture}(300,30)(0,0)
\put(0,0){\parbox{16cm}{\small Figure~6: In the upper panel is
a {\it rough} summation of the {\it sfr}s of the LG dwarf galaxies with 
time (data taken from Mateo \cite{mario}) to obtain the integrated SFH of 
all the LG dwarfs.  The redshifts corresponding to lookback times (for H$_0 
= 50$, q$_0 = 0.5$).  In the middle panel, a wild extrapolation is made; 
the assumption that the integrated SFH of the LG {\it dwarfs} in the upper 
panel is representative of the Universe as a whole.  The resulting star 
formation density of the LG versus redshift is plotted using the same 
scheme as Madau {\it et al.} and Shanks {\it et al.} \cite{sha}, and these 
two models are also plotted and the LG curve is {\bf arbitrarily}, and with 
a very high degree of uncertainty, normalised to the other two models.  In 
the lowest panel the The LG dwarf {\it sfr} as a fraction of the total star 
formation integrated over all time is plotted versus redshift, and the 
Madau curve is also replotted in this form, for the volume of the LG.  This 
highlights the totally different distibution of star formation with 
redshift found from galaxy redshift surveys and what we appear to 
observe in the stellar population of the LG.
}}
\end{picture}
\vskip 3cm

\noindent{the LG.}  dIs represent a small fraction of the total
star formation in the LG. However direct observations of 
the details of the oldest star forming episodes in any galaxy
are limited at best.  There 
are only a few cases for which we have observations which can tell us not 
only when the dominant epoch of star formation has been in a galaxy, but 
also how intense this was.  This is an area where advanced CMD analysis 
techniques have been developed \cite{ts} and telescopes with sufficient 
image quality exist and the required deep, high quality imaging are 
observations are waiting to be made.

Figure~6 summarises what can currently be said
about the SFH of the LG and how 
this compares with the Madau {\it et al.} \cite{madau} and Shanks {\it et 
al.} \cite{sha} redshift survey predictions.  We have not included the 
dominant large galaxies in the LG, the Galaxy and M~31, 
but the SFH of the combined dwarfs is broadly consistent with what is known 
about the SFH of these large systems.  They have, as far as we can tell, had 
a global {\it sfr} that has been gradually but steadily declining since 
their (presumed) formation epoch $>$10~Gyr ago.  There is currently no 
evidence for a particular peak in {\it sfr} around 7$-$9~Gyr ago or any 
other time, as predicted by the Madau-diagram for either large galaxies
or dwarfs. Perhaps if dIs are singled out a population with a star formation
peak in the Madau-diagram range can be found. But at present the statistics
are too limited.  
This is a question 
that would be very useful to study again with HST data and modern CMD 
analysis techniques.  There is 
clearly a total mismatch between the SFH of the LG and the results 
from the redshifts surveys.  This might hint at serious incompleteness 
problems in high redshift galaxy surveys, which appear to miss passively 
evolving systems in favour of small bursting systems.

The recent HST CMD results 
give much cause for optimism that we can hope to sort 
out in detail the SFH of all the different types of galaxies within in the 
LG if only HST would point at them occasionally.  There is also great 
potential for ground based imaging using high quality imaging telescopes 
with large collecting areas, such as VLT is clearly going to be.

\vskip0.3cm
\noindent{{\it Acknowledgments:}} I thank Jay Gallagher, Mario Mateo,
Piero Rosati, Evan Skillman and Alvio Renzini for useful conversations.


%
\vfill
\end{document}